\documentstyle[12pt,epsf]{article}
\font\twlmsy=msbm10 at 12pt
\font\sevenmsy=msbm8
\font\fivemsy=msbm6
\newfam\Bbbfam
\textfont\Bbbfam=\twlmsy
\scriptfont\Bbbfam=\sevenmsy
\scriptscriptfont\Bbbfam=\fivemsy
\def\Bbb{\fam\Bbbfam}

\newcommand{\rf}[1]{(\ref{#1})}
\newcommand{\oh}{\frac{1}{2}}

\newcommand{\bea}{\begin{eqnarray}}
\newcommand{\eea}{\end{eqnarray}}

\newcommand{\beas}{\begin{eqnarray*}}
\newcommand{\eeas}{\end{eqnarray*}}
\newcommand{\beqs}{\begin{displaymath}}
\newcommand{\eeqs}{\end{displaymath}}

\newcommand{\cW}{{\cal W}}

\newcommand{\ben}{\begin{equation}}
\newcommand{\een}{\end{equation}}

\newcommand{\bdm}{\begin{displaymath}}
\newcommand{\edm}{\end{displaymath}}

\newcommand{\pa}{\partial}

\newcommand{\lan}{\langle}
\newcommand{\ran}{\rangle}
\newcommand{\bbR}{{\Bbb R}}

\def\labelmark{}

\newenvironment{formula}[1]{\def\labelname{#1}
\ifx\void\labelname\def\junk{\begin{displaymath}}
\else\def\junk{\begin{equation}\label{\labelname}}\fi\junk}%
{\ifx\void\labelname\def\junk{\end{displaymath}}
\else\def\junk{\end{equation}}\fi\junk\labelmark\def\labelname{}}

{\ifx\void\labelname\def\junk{\end{array}\end{displaymath}}
\else\def\junk{\end{array}\right.\end{equation}}
\fi\junk\labelmark\def\labelname{}\def\junk{}
}

\newcommand{\beq}{\begin{formula}}
\newcommand{\eeq}{\end{formula}}
\newcommand{\beqv}{\begin{formula}{}}

\newcommand{\cWt}{{\widetilde{\cal W}}}
\newcommand{\wt}{\widetilde{w}}
\newcommand{\Nt}{\widetilde{N}}
\begin{document}
\topmargin 0pt
\oddsidemargin 5mm
\headheight 0pt
\topskip 0mm

\addtolength{\baselineskip}{0.20\baselineskip}

\pagestyle{empty}

\begin{flushright}
OUTP 96-38P\\
11th July 1996\\
cond-mat/9607087\\
\end{flushright}
\begin{center}

\vspace{18pt}
{\Large \bf Avalanche size distribution in a \\
random walk model}

\vspace{1 truecm}

%\centerline{}

\vspace{2 truecm}

{\em Thordur Jonsson\footnote{e-mail: thjons@raunvis.hi.is} 
\footnote{Permanent address: University of Iceland,
Dunhaga 3, 107 Reykjavik,
Iceland}}
and
{\em John F. Wheater\footnote{e-mail: jfw@thphys.ox.ac.uk}}

\bigskip

Department of Theoretical Physics, University of Oxford \\
1 Keble Road, Oxford OX1 3NP\\
United Kingdom
\vspace{3 truecm}

\end{center}

\noindent
{\bf Abstract.} We introduce a simple model for
the size distribution of avalanches based on the idea that the 
front of an avalanche can be described by a directed random walk.
The model captures some of the qualitative features of
earthquakes, avalanches and other self-organized critical phenomena 
in one dimension.
We find scaling laws relating the frequency, size and 
width of avalanches and
an exponent $4/3$ in the size distribution
law.

\vfill

\newpage
\pagestyle{plain}

\section{Introduction}

Driven dissipative systems arise in modelling many different phenomena
in Nature, e.g.\ avalanches, forest fires, earthquakes and 
Darwinian evolution.  Such systems
frequently exhibit long range temporal and spatial correlations and
scaling laws analogous to statistical mechanical systems at a critical
point.  The concept of self-organized criticality was introduced \cite{bak1} 
to describe systems where critical behaviour arises without the
fine tuning of any parameter.

The study of self-organized criticality is essentially a study of a
family of dynamical systems which evolve in time in small or large
steps which have a power law distribution and 
we shall call ``avalanches'' but they may have a different
interpretation.  The canonical example of a self-organized critical system is
the abelian sandpile model \cite{bak1,dhar} 
where an integer variable $z_i$ is assigned to each
site $i$ in a lattice and the dynamics is given by choosing a random site
$i$ in the lattice, increasing $z_i$ by 1, $z_i\to z_i+1$, and if
$z_i$ exceeds a threshold value $z_c$, then an avalanche start and the
value of $z_i$ is decreased by redistributing some of $z_i$ among the
neighbours.  This in turn may push some of the $z_j$ over threshold,
where $j$ is a neighbour of $i$, and so on.  These rules are then
supplemented by appropriate boundary conditions.  See \cite{christensen} 
for a recent detailed discussion.

It is natural to introduce a mean field theory for self-organized
criticality \cite{mft1}.  This has been done in many different ways
for different models yielding identical critical exponents
 \cite{mft2,mft3,flyvbjerg,zapperi}.   
One of the approaches is to
study sandpiles on a Bethe lattice \cite{mft2}.
In this case different parts of an avalanche front are uncorrelated
and one obtains the scaling law
\beq{0}
N_w(k)\sim k^{-3/2}
\eeq
where $N_w(k)$ is the frequency of avalanches involving $k$ sites.
We shall refer to $k$ as the {\em width} of the valanche.

In this paper we introduce a simple model for the propagation of a
one-dimensional 
avalanche front.  This model was suggested by studying earthquakes in
the Burridge--Knopoff model \cite{burridge,cl1,cl4,nakanishi,jonsson}.
The basic 
idea is to assume that
neighbouring parts of an avalanche are correlated in the following
way.  If we label elements of the avalanche by integers $i$ and an
element $i$ moves a distance $h_i$ then the
displacement of its neighbour, 
labelled by $i+1$, is distributed with a probability
distribution $P_i(h_{i+1})=\phi (h_{i+1}-h_i)$ which is 
centered on $h_i$ but otherwise independent of $i$.  This distribution is
modified by an appropriate boundary condition at $h_{i+1}=0$ amounting
to a certain killing probability for the avalanche.  In this paper we
shall in fact work with the simplest possible probability distribution
$\phi$ which corresponds to the avalanche front performing a Bernoulli
random walk on the positive integers and terminating once it returns
to $0$.  In this case one can perform explicit calculations and
we find an exponent $3/2$ in the scaling law relating the frequency of
avalanches to their width.

If we define the
{\em size} of an avalanche as
\beq{01}
A=\sum_ih_i,
\eeq
then we find that the frequency of avalanches of size $A$ is given by 
\beq{1}
N(A)\sim A^{-4/3}.
\eeq
Sometimes, e.g.\ in earthquake models, it is
natural to place an upper bound on the maximal allowed displacement $h_i$ 
and in that case the power law turns into an exponential decay for
large $A$.  The relation to earthquakes 
is discussed in more detail in \cite{quake}.

\section{The random walk model}
In this section we define the random walk model.  We introduce a class
of random walks (or paths) and each path will correspond to an avalanche of a
particular form.  Questions about the size, shape and frequency of
avalanches can therefore be translated into questions about the number
of paths satisfying the appropriate conditions.  In this model we are
not able to address questions about the correlations between
different avalanches nor discuss the self-organized critical 
spatial structure that arises in realistic models.
  
We shall refer to the discrete elements participating in an
avalanche as blocks.  
%
%
%In this section we recall the definition of the BK-model and the main
%features of its behaviour.   The homogenous one-dimensional 
%BK-model is a linear chain of blocks of equal masses each of which is
%connected to its two nearest neighbours by springs with a spring
%constant $k_c$.  The blocks slide on a surface subject to a velocity
%dependent friction force $F$ which decreases with velocity.  This
%surface may be thought of as one side of the earthquake fault.  In
%addition each block is connected by another spring with spring
%constant $k_p$ to a surface which moves with a constant velocity $v$
%and shouldd be thought of as the other side of the earthquake fault.
%etc. etc.
%etc.
%
Consider a semi-infinite chain of blocks $M_i$, 
lying along the $x$-axis,
labelled by the
non-negative integers which can be regarded as the $x$ coordinates of
the blocks.  We assume that the blocks move in integer steps in the
direction of the $y$ axis with the $x$ coordinate unchanged.
In applications to many one-dimensional systems, e.g.\ earthquakes and
sandpiles, the $x$ coordinate would be the $x$ coordinate of the
blocks at time $0$ while the $y$ coordinate, in our present terminology,
would be the addition to the $x$ coordinate in an avalanche.

We imagine that the
block labelled by $i=1$ is the first one to move in an avalanche and
the next block to move is $M_2$ etc.  For convenience we assume that
the block at the origin $M_0$ is fixed at all times, but this
assumption is not at all essential.  We could easily have another
avalanche moving to the left and this second avalanche would be
uncorrelated to the one moving to the right.  We assume that
$M_1$ moves a distance 1 along the $y$-axis.  This is the initial
condition for an avalanche.  We now assume that $M_2$
moves a distance 2 along the $y$-axis with probability $\oh$ and a
distance $0$ with probability $\oh$.  In general, given that the block
$M_n$ has moved a distance $h_n$, block $M_{n+1}$ moves a distance
$h_n\pm 1$ with equal probabilities.  By definition, the avalanche
stops once the next block to move in fact does not move.  The minimal
avalanche is the one where $M_2$ does not move and this is the most
likely event with probability $\oh$.  It is sometimes natural to place
an upper bound $h$ on how far the blocks can move so we modify the rules
 by requiring $M_{n+1}$ to move a distance $h-1$, if $M_n$
has moved a distance $h$.  Of course it would be natural to 
allow neighbouring
blocks to move the same distance.  This would not affect any of the
conclusions but make some of our equations more complicated. 
In this formulation an avalanche can be regarded as a directed random
walk on the non-negative integers which is reflected from $h$ and
stops when it returns to $0$.  

We can write 
\beq{5}
h_n=\sum _{i=1}^n \sigma _i,
\eeq
where $\sigma _1=1$ and $\sigma _i$ for $i\geq 2$ 
is a random variable which takes
the values $\pm 1$ with equal probabilities.  The duration $T$ of the
walk is defined by
\beq{6}
h_T=0, ~~~h_n>0,~~0<n<T.
\eeq
The size (moment in seismology) $A$ of the avalanche 
is given by the area under the graph of
the function $h_n$, i.e.
\beq{7}
A=\sum _{n=1}^Th_n.
\eeq
We are interested in calculating the frequency of avalanches with
size $A$ and the relation between $A$ and the width
$T$.  

\section{Short time behaviour}
We begin by considering the model in the absence of an upper
cutoff $h$, i.e.\ we put $h=\infty$ and return to the case of finite $h$ later.
A directed random walk (or path) 
is one where the $x$ coordinate increases by 1
in each step.
Let $\cW$ denote the set of all directed
walks in the positive quadrant of the $xy$-plane which start from $(0,0)$
and return to the $x$-axis. Such a walk is located at $(1,1)$ after
one step.  Let $N(A,T)$ denote the number of paths in
$\cW$ which return to the $x$-axis after $T$ steps and
whose graph, together with the $x$ axis, 
encloses an area $A$ given by Eq.\ \rf{7}, see Fig.\ 1.  Let
\beq{8}
N(T)=\sum _AN(A,T).
\eeq  
The probability that the walk returns to the origin 
after $T$ steps is given by
\beq{9}
P(T)=2^{-T+1}N(T).
\eeq
We divide by $2^{T-1}$ since the first step in the walk is given.
Similarly, the conditional probability $P(A|T)$ 
that a walk covers an area $A$, given that
it lasts a time $T$, can be written as
\beq{10}
P(A|T)={N(A,T)\over N(T)}.
\eeq
It follows that the probability that an avalanche has size $A$ is
given by 
\beq{11}
P(A)=2^{-T+1}\sum _TN(A,T).
\eeq	
It is well known from the solution of the classical gambler's ruin
problem, see e.g.\ \cite{feller}, that
\beq{12}
P(T)={2^{-T+1}\over T-1}\left(\begin{array}{cc}T-1\\T/2
\end{array}\right)
\eeq
if $T$ is even and $0$ otherwise. 
For large $T$
\beq{13}
P(T)\sim T^{-{3\over 2}}.
\eeq
Since we are discussing random walks it is natural to expect the 
average height $\lan h_n\ran$ of a walk which lasts a time $T$ to scale like
$\sqrt{T}$ for large $T$.  The average area 
$A$ should therefore grow as $T^{3/2}$ for
such walks.   A priori one would expect entropic repulsion to play 
a role here so $3/2$ should be a lower bound on the ``average area exponent'' 
but, as demonstrated below, there is in fact no
shift  away from the naive value of this exponent. 
   
If we consider directed random walks that are allowed to cross the
$x$ axis and define the area under the walks to be positive if the walk is in
the upper half plane and negative when it is in the lower half plane,
we can use \rf{5} and \rf{7} to express the area as a sum of
independent but non-identical random variables.  The generalized
central limit theorem \cite{feller} applies to this sum and we find
that asymptotically the area is normally distributed around zero with a
variance $T^3$.  
Assuming that $P(A|T)$ is normally distributed around its
average value with a variance $T^3$ we expect that for large $A$ we have
\bea
P(A)&=&\sum _TP(A|T)P(T)\nonumber\\
    &\sim &\int_0^{\infty}T^{-3}\exp\left({(A-T^{3/2})^2\over
T^3}\right)\, dT\nonumber\\  
    &\sim & A^{-4/3}.\label{X1}
\eea
Below we shall verify that the asymptotic behaviour of $P(A)$ is
indeed given by Eq.\ \rf{X1} even though one can prove that 
$P(A|T)$ is in fact not normally distributed by computing its first
few moments.

Let $\cWt$ denote the class of directed walks in $\cW$
which avoid the line $y=1$ until they return to $y=0$, i.e.\ if
$w\in\cWt$ and $w$ returns at time $T$ then $w (x)>1$
for $1<x<T-1$, where
$w (t)$ denotes the
$y$-coordinate of the path for $x=t$.  Let us denote
by $\Nt(A,T)$ the number of paths in $\cWt$ which return to $0$ at time $T$  
and cover an area $A$.  Then
\beq{14}
\Nt(A,T)=N(A-T+1, T-2),
\eeq
see Fig.\ 2.
Now consider any directed 
walk $w \in \cW$ which lasts a time $T>2$ and covers an area $A$.  
  Let $T_1$ denote the smallest
integer $>1$ such that $w (T_1)=1$.  The largest possible value of
$T_1$ is of course $T_1=T-1$.  If we cut the path $w$ in two
pieces at the point $(T_1,1)$ then we can associate uniquely to $w$ 
two paths, $\wt \in\cWt$ and $w _1\in \cW$, of duration
$T_1+1$ and $T-T_1+1$, respectively, see Fig.\ 3.  In the extreme case
$T_1=T-1$ the second walk is the trivial one of length 2.  
If we denote the area
under the first walk by $A_1$ then the area under the second one 
equals $A-A_1+1$ and we find that
\bea
\lefteqn{N(A,T)=\delta _{A1}\delta _{T2}+\sum_{T_1=1}^{T-1}
\sum_{A_1=1}^{A}\Nt(A_1,T_1+1)N(A-A_1+1,T-T_1+1)}\nonumber
\\
&=&\delta_{A1}\delta_{T2}+\sum_{T_1=1}^{T-1}\sum_{A_1=1}^{A}N(A_1-T_1,T_1-1)
N(A-A_1+1,T-T_1+1),\label{15}
\eea
by Eq.\ \rf{14}.  The first term on the right side of Eq.\ \rf{15}
corresponds to the two step path.  
We define the generating function $f(z,u)$ for the
numbers $N(A,T)$ by
\beq{16}
f(z,u)=\sum_{T=2}^\infty\sum_{A=1}^\infty N(A,T)z^Au^T,
\eeq
which is convergent for $|z|<1$ and $|u|\leq \oh$.  Eq.\  
\rf{15} can now be rewritten as
\beq{17}
f(z,u)=zu^2+f(z,u)f(z,uz).
\eeq
For $z=1$ we can easily solve this equation and find
\beq{18}
f(1,u)=\oh -\oh\sqrt{1-4u^2},
\eeq
which for $u=\oh$ takes the value $\oh$ in accordance with Eq.\
\rf{11}.

For general values of $z$ and $u$ Eq.\ \rf{17} is not explicitly
soluble.  We note however that the equation can be rearranged to read
\beq{19}
f(z,u)={zu^2\over 1-f(z,uz)},
\eeq
which, upon iteration, yields a continued fraction expansion\footnote{
It is perhaps of interest to note that the 
continued fraction \rf{20} appeared in a letter from Ramanujan to Hardy
written in 1913 \cite{hardy} 
where it was used to express some remarkable identities,
one of which can be written, in our notation, 
$$
f(e^{-\pi},i e^{\pi /2})=-\left( \sqrt{{5+\sqrt{5}\over
2}}-{\sqrt{5}+1\over 2} \right) e^{2\pi /5}.
$$
However, the Ramanujan identities seem unrelated to the statistical
properties of the random walk model we are interested in.}
 for $f(z,u)$
\beq{20}
f(z,u)={zu^2\over 1-{\displaystyle z^3u^2\over \displaystyle 1 -{
\displaystyle z^5u^2\over \displaystyle 1-\ddots }}}.
\eeq
%For $u=\oh$ his function is known \cite{paf} to have a natural
%boundary at $|z|=1$ meaning that it cannot be analytically continued
%beyond the unit circle.  

Looking at Eq.\ \rf{X1} we expect the average size of an avalanche
to diverge, i.e.\ we expect
\beq{21x}
\lim _{u\uparrow \oh}\left.{\pa \over \pa z}f(z,u)\right|_{z=1}=\infty .
\eeq
Indeed, we shall prove that 
\beq{22}
\left.{\pa ^n\over \pa z^n}f(z,u)\right|_{z=1}\sim (1-4u^2)^{\oh -{3n\over
2}}
\eeq
as $u\uparrow \oh$, for any $n\geq 1$.  Let us define
\beq{23}
P_u(A)=\sum _Tu^{T-1}N(A,T).
\eeq
Then
\beq{24}
\sum _AA^nP_u(A)=\left.\left(z{\pa \over \pa z}\right)^nf(z,u)\right|_{z=1}.
\eeq
For any path of duration $T$, covering an area $A$, it is easy to
see that
\beq{24x}
{3\over 2}T-2\leq A\leq {1\over 4}T^2.
\eeq
The lower bound is obtained by considering the path which zigzags
between 1 and 2 and covers the smallest possible area while the upper
bound corresponds to the triangular path that climbs to height $T/2$
in time $T/2$ and then descends to zero in time $T/2$. 
It follows that for $u$ smaller than but close to $\oh$ we have the
bounds
\beq{24y}
e^{-c_1(1-2u)A}P(A)\leq P_u(A)\leq e^{-c_2(1-2u)\sqrt{A}}P(A)
\eeq
where $c_1$ and $c_2$ are positive constants.  It is natural to regard
$u$ as a temperature-like parameter and $u=\oh$ as a critical point so
we expect scaling as this point is approached.  
Assuming that 
\beq{25}
P_u(A)\sim A^{-\gamma}F(\sqrt{1-4u^2}A^{\beta})
\eeq
for large $A$, where $F$ is a function decaying more rapidly than any
power, we find that
\beq{26}
\sum _AA^nP_u(A)\sim (1-4u^2)^{{\gamma -n-1\over 2\beta}}.
\eeq
Since this is valid for any $n$ it follows from Eq.\ \rf{22} that
\beq{27}
\beta ={1\over 3}~~~{\rm and}~~~\gamma={4\over 3}.
\eeq
This is of course consistent with the inequalities \rf{24y} which imply
that ${1\over 4}\leq \beta\leq\oh$.

In order to verify Eq.\ \rf{22} let us denote the derivatives of the
generating function $f$ with respect to the first and second argument
by $\pa _1f$ and $\pa _2f$, respectively.  We begin by considering the
case $n=1$.  Differentiating Eq.\ \rf{17} with respect to $z$ we obtain
\beq{28}
\pa _1f(z,u)=u^2+f(z,uz)\pa
_1f(z,u)+f(z,u)(\pa_1f(z,uz)+u\pa_2f(z,uz)).
\eeq
Putting $z=1$, rearranging and using Eq.\ \rf{18} we obtain
\bea
\pa _1f(1,u) &=& {u^2+u\pa _2f(1,u)\over 1-2f(1,u)}\nonumber\\
             &=&{u^2+2u^2(1-4u^2)^{-\oh}\over\sqrt{1-4u^2}}\nonumber\\
             &\sim&{1\over 1-4u^2}.
\eea
Assume now that Eq.\ \rf{22} holds for $n\leq N-1$ where $N\geq 2$.
Differentiating Eq.\ \rf{17} $N$ times with respect to $z$ yields
\beq{999}
\pa _1^{N}f(1,u)=\sum _{k=0}^{N}
\left(\begin{array}{cc}N\\ k\end{array}\right)
\pa_1^kf(1,u)
\sum_{j=0}^{N-k}\left(\begin{array}{cc}N-k\\
j\end{array}\right)u^{N-j-k}\pa_1^j\pa_2^{N-j-k}f(1,u).
\eeq
Rearranging we find that
\bea
\lefteqn{\pa _1^Nf(1,u)=
{1\over 1-2f(1,u)}\sum_{k=0}^{N-1}
\left(\begin{array}{cc}N\\ k\end{array}\right)
\pa _1^kf(1,u)}&&\nonumber\\&&
\times \sum _{j=0,j\neq N}^{N-k}\left(\begin{array}{cc}N-k\\
j\end{array}\right)u^{N-j-k}\pa _1^j\pa _2^{N-j-k}f(1,u).\label{66}
\eea
By the inductive hypothesis the most singular terms on the right hand side
of Eq.\ \rf{66} correspond to $j+k=N$ if $k>0$ and $j=N-1$ when $k=0$.
The desired result follows.

Working slightly harder we can determine the coefficient of the
leading divergence of the $N$th moment of $P_u(A)$ and this allows us
to place a further restriction on the function $F$ introduced above.  
In view of  
Eq.\ \rf{22} one can write
\beq{67}
\pa_1^Nf(1,u)=C_N(1-4u^2)^{\oh -{3N\over 2}}+O((1-4u^2)^{1-{3N\over
2}}).
\eeq
It is straightforward to check that Eq.\ \rf{66} determines the
following recursion relation for the coefficients $C_N$
\beq{68} 
C_N=\sum_{k=1}^{N-1}\left(\begin{array}{cc}N\\
k\end{array}\right)C_kC_{N-k}+(3N-4)NC_{N-1}.
\eeq
It follows that up to power corrections 
\beq{7777}
C_N\sim (2N)!
\eeq
for large $N$.   

Suppose now that the function $F$ in \rf{25} 
is an exponential function of a power, i.e.\
\beq{69}
F(x)= e^{-\alpha x^q}
\eeq
for some constants $\alpha, q>0$.  Using the ansatz 
\rf{25} to calculate the $N$th moment of the area distribution 
we find that $q$ is fixed to equal $3/2$  by the asymptotic formula 
\rf{7777}.  We therefore expect that
\beq{70} 
P_u(A)\sim A^{-4/3}e^{-\alpha (1-4u^2)^{3/4}A^{1/2}}
\eeq
in agreement with the bound \rf{24y}.

This completes our disussion of the frequency-size distribution of
avalanches where we do not need to take the upper cutoff $h$ into
account, i.e.\ the exponent $4/3$ governs the size distribution of
{\em small} avalanches in the presence of a cutoff.  
We now turn to the study of the tail of the
size distribution and will see that for large $A$ the probability
$P(A)$ falls off exponentially when the upper bound $h$ is taken into
account.  We also estimate the area for which we have a transition
from a power law to an exponential decay.

\section{Long time behaviour}
We now consider a directed random walk with a reflecting barrier at
height $y=h$.  Let $p_i(t)$ be the probability of the walk being at
height $i$ after $t$ steps.  Then the initial condition is 
\beq{71}
p_i(1)=\delta _{i1}.
\eeq
We let $p(t)$ denote the column vector whose $i$th entry is $p_i(t)$.
Then
\beq{72}
p(t)=\left(\oh M\right)^{t-1}p(1)
\eeq
where $M$ is the matrix
\beq{73}
M=\left(\begin{array}{ccccccc}0&1&0&0&\ldots&0&0\\
2&0&1&0&0&\cdots&0\\
0&1&0&1&0&\cdots&0\\
0&0&1&0&1&\cdots &0\\
\vdots&\vdots&&&\ddots&1 &\vdots\\
0&\cdots & & 0&1&0&0\\
0&\cdots & & & 0&1&0
\end{array}\right).
\eeq
The probability that a walk lasts exactly $T$ steps is evidently
\beq{74}
P(T)=\oh p_1(T-1).
\eeq
Let $\lan \cdot ,\cdot \ran$ denote the standard inner product on $\bbR
^{h+1}$ and $e_i$, $i=0,\ldots ,h$, the standard orthonormal basis.
Then we can write
\beq{75}
P(T)=\lan e_0,\left(\oh M\right)^{T-1}e_1\ran .
\eeq
Let $D$ denote the matrix whose elements are defined by
\beq{76}
D_{ij}=z^i\delta_{ij}
\eeq
$i,j=0,\ldots ,h$.  If $P(A,T)$ denotes the probability that a walk
lasts a time $T$ and covers an area $A$ then $P(A,T)$ is given by the
coefficient of $z^A$ in the matrix element
\beq{77}
\lan e_0,\left(MD\right)^{T-1}e_1\ran .
\eeq
The generating function for the probabilities $P(A,T)$ can therefore
be expressed as
\beq{78}
Q(z,u)=\lan e_0,{u^2MD\over 1-uMD}e_1\ran ,
\eeq
since the Neumann series for the inverse of $(1-uMD)$ is easily seen
to converge for $|u|\leq \oh$ and $|z|\leq 1$. 

Eq.\ \rf{78} allows us in principle to calculate $P(A,T)$.  However,
the interesting feature of $P(A,T)$ is that it falls exponentially
with $T$ and $P(A,T)=0$ unless $A\leq hT$.  It follows that
$P(A)=\sum_TP(A,T)$ falls exponentially with $A$ provided $A$ is large
enough.  In order to establish this exponential decay it suffices to show
that $Q(1,u)$ is finite for some $u>\oh$.  We can write
\beq{79}
Q(1,u)=\oh\lan e_0,{1\over \lambda -\oh M}e_1\ran
\eeq
where $\lambda =(2u)^{-1}$.  Evaluating the matrix element in Eq.\
\rf{79} is an elementary calculation and we find
\beq{80}
Q(1,u)={1\over
4}{\cos\left((h-2)\theta\right)\over\cos\left((h-1)\theta\right)}
\eeq
where
\beq{81}
e^{i\theta}= \lambda +i\sqrt{1-\lambda ^2}
\eeq
and we are assuming $\lambda \leq 1$.  The first singularity of
$Q(1,u)$ as $u$ moves beyond $\oh$ is encountered for the smallest
$\theta \in [0,2\pi )$ for which the denominator in Eq.\ \rf{80}
vanishes, i.e.
\beq{82}
\theta ={\pi\over 2(h-1)}.
\eeq
It follows that the radius of convergence of $Q(1,u)$ is
\beq{83}
r=\oh\sqrt{1+\tan ^2{\pi\over 2(h-1)}},
\eeq  
and for large $A$ we find
\beq{84}
P(A)\leq Ce^{-cA/h^2}
\eeq
where $c$ and $C$ are positive constants and $c$ can be taken to be
independent of $h$.

The exponential decay of $P(A)$ takes over from the power
law found in the previous section for $A\approx h^3$ since a random
walk must have at least $h^2$ steps in order to feel the effect of the
reflecting barrier at $y=h$.  In order to prove this note that we can
write
\beq{85}
P(T)={2^{1-T}\over 2\pi i}\oint {Q(1,u)\over u^{T+1}}du,
\eeq
where the contour encloses the unit disc in the complex plane.
Calculating the residues we find that
\beq{86}
P(T)=2h^{-1}\sum _{n=0}^{2(h-1)} \sin ^2{\pi (1+2n)\over h-1}\cos
^{T-1}{\pi (1+2n)\over h-1}
\eeq
and consideration of this formula at large $h$ shows that
$P(T)$ crosses over from exponential decay to a $T^{-3/2}$ decay
for $T\approx h^2$.

\section{Discussion}
By universality, we do not expect the principal results we have
obtained to change if we replace the simple Bernoulli random walk, 
considered in this paper, by a random walk with any rapidly decaying
transition function.  The exponent value $4/3$ ought to be
universal as will be exponential decay for large $A$ in the presence
of an upper cutoff $h$.

We do, however, expect to be able to change the power law decay by
considering strongly correlated random walks.  In earthquake models, for
example, the size distribution exponent for small and intermediate
size events varies from around $2/3$ to
values greater than $1$.  It is clear that by looking at the
statistics of avalanches in a realistic system one can concoct a
random walk model with an identical avalanche distribution.

The more interesting problem of understanding correlations between
different avalanches cannot be studied in the random walk framework
unless one introduces different interacting random walks.
The principal virtue of the model we have discussed is that it gives
us a qualitative and quantitative 
insight into the genesis of power law distribution
for avalanches without introducing any complicated dynamics.

\newpage

\noindent
{\bf Fig. 1}  A directed random walk of duration $T=12$.

\epsfxsize\textwidth\epsfbox{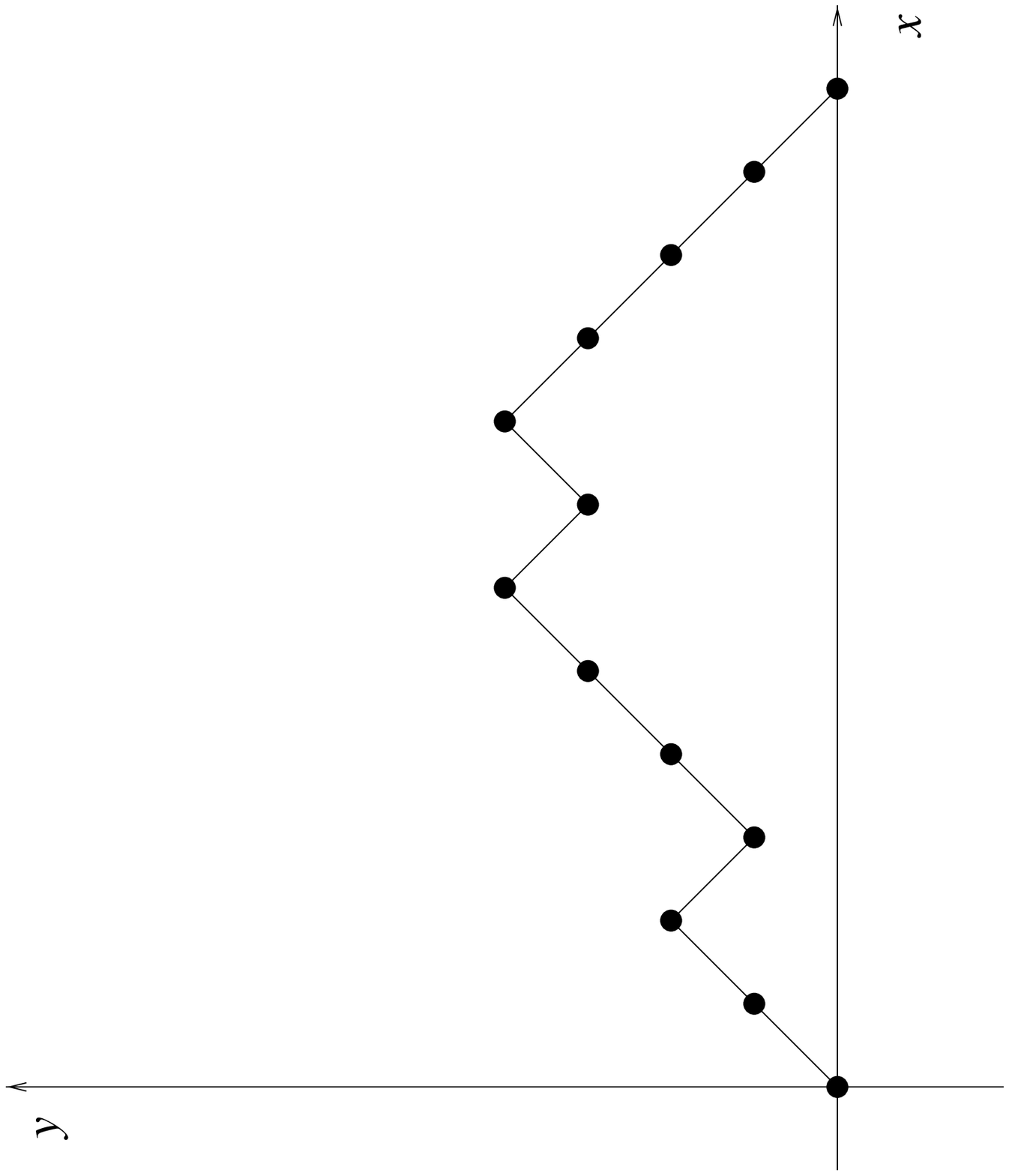}
\newpage

\noindent
{\bf Fig. 2.}  This figure illustrates the one to one correspondence
between paths in $\cW '$ of duration $T$ and paths in $\cW$ of duration
$T-2$.

\epsfxsize\textwidth\epsfbox{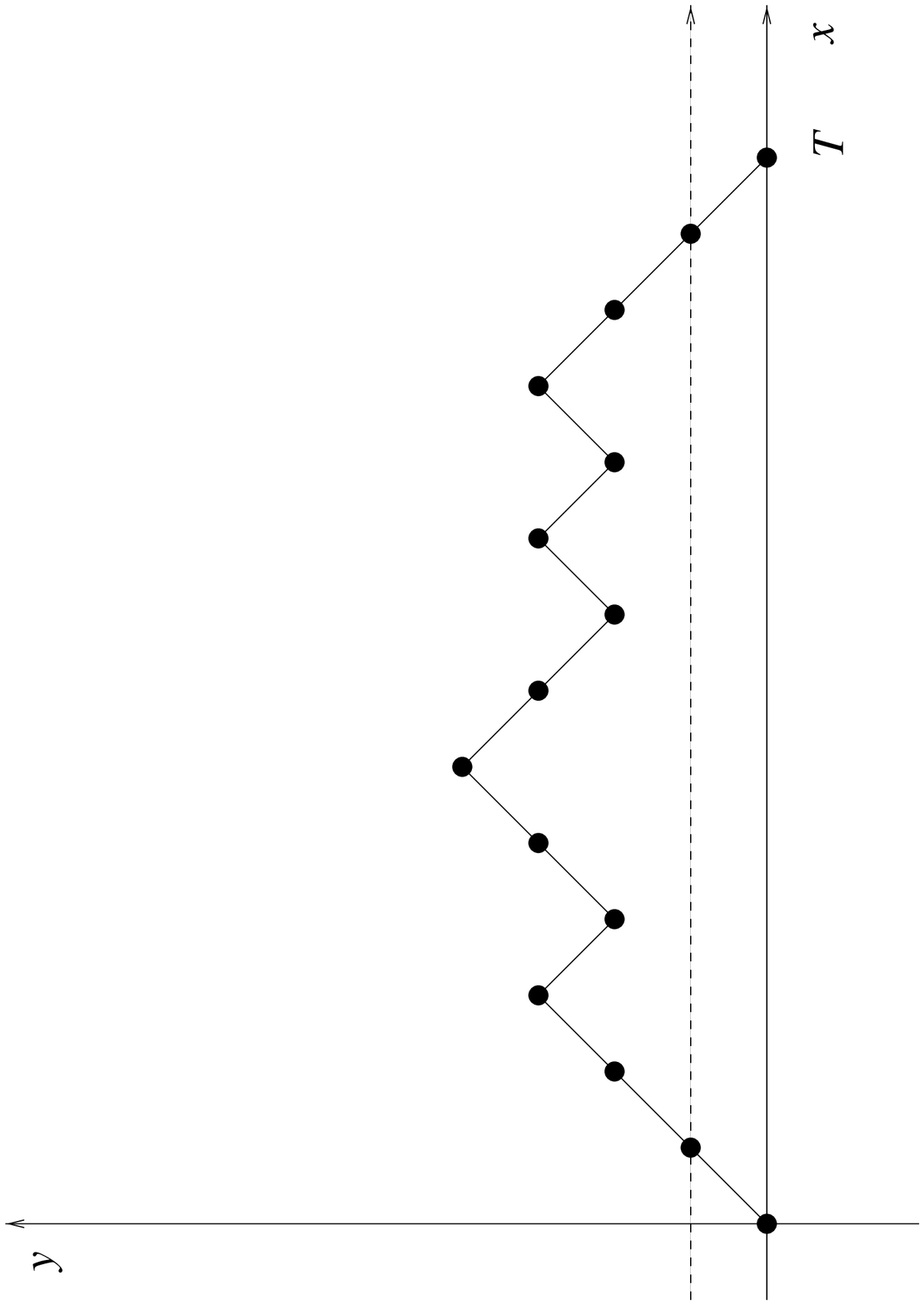}
\newpage

\noindent
{\bf Fig. 3.} This figure illustrates how one can uniquely decompose any
directed path into a pair of paths in $\cW '$ and $\cW$.

 \epsfxsize\textwidth\epsfbox{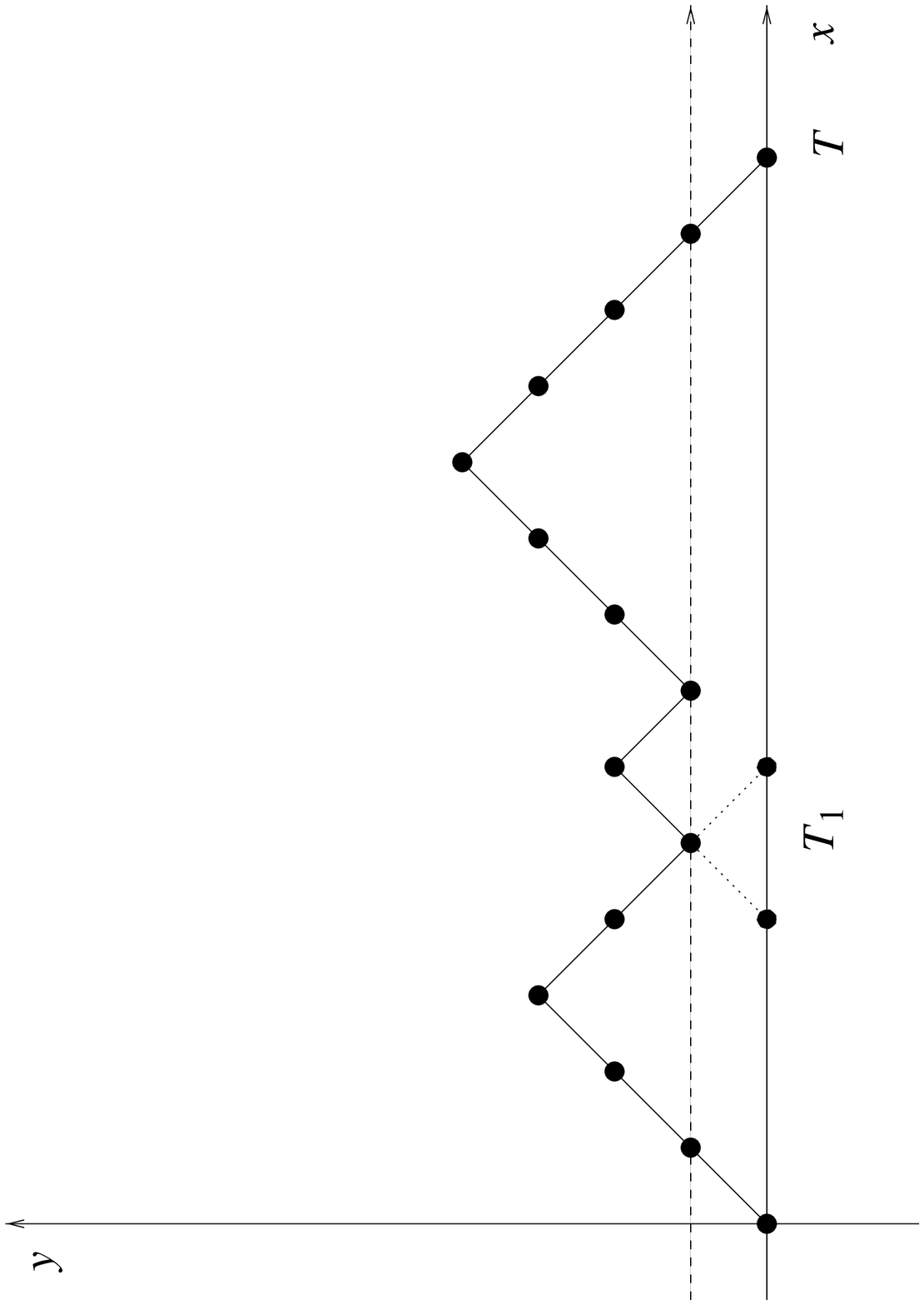}


\begin{thebibliography}{99}
\bibitem{bak1}P. Bak, C. Tang and K. Wiesenfeld, Phys. Rev. Lett. 59 (1987) 
381.
\bibitem{dhar}D. Dhar, Phys. Rev. Lett. 64 (1990) 1613.
\bibitem{christensen}K. Christensen and Z. Olami, Phys. Rev. E 48 (1993) 3361. 
\bibitem{mft1}C. Tang and P. Bak,  {\em J. Stat. Phys.} {\bf 51} (1988) 797 - 802.
\bibitem{mft2}D. Dhar and S. N. Majumdar,  {\em J. Phys. A: Math.. Gen.} {\bf 23} (1990)
4333. 
\bibitem{mft3}S. A. Janowsky and C. A. Laberge, {\em J. Phys. A: Math. Gen.} {\bf 26}
(1993) L973.
\bibitem{flyvbjerg}H. Flyvbjerg, K. Sneppen and P. Bak, Phys. Rev. Lett. 71 (1993) 4087.
\bibitem{zapperi}S. Zapperi, K.~B. Lauritsen and H.~E. Stanley,
Phys. Rev. Lett. 75 (1995) 4071.
\bibitem{burridge} R. Burridge and L. Knopoff, 
{\em Bull. Seismol. Soc. Am.} {\bf 57} (1967) 341.
\bibitem{cl1} J. M. Carlson and J. S Langer,  {\em Phys. Rev. A} {\bf 40} (1989) 6470.
\bibitem{cl4}J. M. Carlson, J. S. Langer and B. E. Shaw, 
 {\em Rev. Mod. Phys.} {\bf 66} (1994) 657.
\bibitem{nakanishi} H. Nakanishi,  {\em Phys. Rev. A} {\bf 41} 
(1990) 7086.
\bibitem{jonsson}T. Jonsson and S. F. Marinosson,  {\em
Physics Letters A} {\bf 207} (1995) 165.
\bibitem{gutenberg}B. Gutenberg and C. F. Richter, {\em Seismicity of
the Earth}, Princeton University Press, Princeton, 1954.
\bibitem{quake}T. Jonsson and J.~F. Wheater, Gutenberg--Richter law in
a random walk model for earthquakes, Oxford University preprint, OUTP 9639P.
\bibitem{feller}W. Feller, {\em An introduction to probability theory
and its applications, Vol. I}, John Wiley and Sons, New York, 1968.
\bibitem{hardy}G. H. Hardy, {\em Ramanujan}, Cambridge University Press, 1940.


\end{thebibliography}
\end{document}